\documentclass{article}
\usepackage{booktabs}
\usepackage{geometry}
\usepackage{siunitx}
\usepackage[table]{xcolor}
\usepackage{multirow}
\usepackage{graphicx}
\graphicspath{{./Figures/}}
\usepackage{colortbl}
\usepackage[table]{xcolor}
\usepackage{float}
\usepackage{multirow}
\usepackage{longtable}
\usepackage{subfig}
\usepackage{adjustbox}
\usepackage{array,makecell}
\usepackage{enumitem} 
\usepackage{mathtools}
\usepackage{mathdots}
\usepackage{changepage}
\usepackage{booktabs}
\usepackage{tcolorbox}
\usepackage{hyperref}
\usepackage{filecontents}
\usepackage{tikz}
\usepackage{tkz-graph}
\usepackage{chemfig}
\usepackage[title]{appendix}
\usepackage{pgfplots}
\pgfplotsset{compat=1.17}
\usepackage{lscape}
\usepackage{subcaption}
\usepackage{arydshln}
\usepackage{pgfplots}
\pgfplotsset{compat=1.18}
\usepackage{placeins}
\usepackage{longtable}
\usepackage{booktabs}
\usepackage[table]{xcolor}

\definecolor{lightgray}{gray}{0.92}
\newcommand{\zero}{\textit{0}}
\usepackage{amsmath, amsthm, amsfonts, amssymb}

\theoremstyle{definition}

\theoremstyle{remark}

\title{Linear to Neural Networks Regression: QSPR of Drugs via Degree-Distance Indices }

\author{ M. J. Nadjafi-Arani$^{a,b}$, S. Sorgun$^{a,*}$, M. Mirzargar$^{a,b}$  \\
	\small $^{a}$Department of Mathematics, Nevşehir Hacı Bektaş Veli University, 50300, Nevşehir, TÜRKİYE \\
	\small $^{b}$ Faculty of Science, Mahallat Institute of Higher Education, Mahallat, I. R. Iran \\
	\small $^{*}$Corresponding author \tt{srgnrz@gmail.com;szrsorgun@gmail.com} \\
	\small \tt{mjnajafiarani@mahallat.ac.ir; mjnajafiarani@gmail.com} \\
        \small \tt{m.mirzargar@mahallat.ac.ir } \\
}	
\begin{document}
\let\cleardoublepage\clearpage
	
	\maketitle	
	\begin{abstract}
		\noindent
		
		This study conducts a Quantitative Structure–Property Relationship (QSPR) analysis to explore the correlation between the physical properties of drug molecules and their topological indices using machine learning techniques. While prior studies in drug design have focused on degree-based topological indices, this work analyzes a dataset of 166 drug molecules by computing degree-distance-based topological indices, incorporating vertex-edge weightings with respect to different six atomic properties (atomic number, atomic radius, atomic mass, density, electronegativity, ionization).  Both linear models (Linear Regression, Lasso, and Ridge Regression) and nonlinear approaches (Random Forest, XGBoost, and Neural Networks) were employed to predict molecular properties. The results demonstrate the effectiveness of these indices in predicting specific physicochemical properties and underscore the practical relevance of computational methods in molecular property estimation. The study provides an innovative perspective on integrating topological indices with machine learning to enhance predictive accuracy, highlighting their potential application in drug discovery and development processes. This predictive may also explain that establishing a reliable relationship between topological indices and physical properties enables chemists to gain preliminary insights into molecular behavior before conducting experimental analyses, thereby optimizing resource utilization in cheminformatics research.

	\end{abstract}
	
	\section{Introduction}
		Studies on finding relationships between the physical properties of molecules and their topological indices are frequently found in the literature. In particular, QSPR analyses explaining such relationships have been extensively studied in drug design research [\cite{alz}-\cite{viti}]. The first QSPR study on vertex-edge weighted molecular graphs was conducted in \cite{sezer}. The mentioned studies have utilized degree-based topological indices. In these studies, the topological indices of molecular graphs of drugs have been calculated, and some regression models have generally been used to relate them to physical properties. 

    In recent years, machine learning (ML) techniques have been extensively employed in chemistry to predict physicochemical properties, particularly when experimental data are limited \cite{huang,qi}. Studies such as \cite{qi} have demonstrated the effectiveness of ML approaches in handling small datasets by leveraging appropriate feature representations. Inspired by these advancements, we applied both linear and non-linear regression models to explore the relationship between physicochemical properties and topological indices. Specifically, linear regression, Lasso, and Ridge regression were utilized to capture simple linear dependencies, while Random Forest, XGBoost, and Neural Networks were employed to model more complex, non-linear interactions.

Our findings underscore the importance of selecting appropriate regression techniques based on the characteristics of the dataset, as different models demonstrated varying predictive accuracy across various molecular properties. Linear regression methods provided interpretable relationships between topological indices and physicochemical properties, while nonlinear approaches, such as Random Forest, XGBoost, and Neural Networks, exhibited superior predictive performance by capturing complex dependencies within the data.

Topological indices are derived through mathematical computations on the molecular graph, representing structural features of a molecule without requiring experimental measurements. In contrast, determining physicochemical properties often necessitates costly and time-consuming laboratory procedures. Establishing a reliable relationship between topological indices and physical properties enables chemists to gain preliminary insights into molecular behavior before conducting experimental analyses, thereby optimizing resource utilization in cheminformatics research.

In this study, we conducted a Quantitative Structure–Property Relationship (QSPR) analysis of some drugs using machine learning techniques. A dataset of 166 drug molecules was analyzed, and their vertex-edge weightings were used to compute degree-distance-based topological indices. The results not only highlight the correlation between these indices and physical properties but also identify which molecular properties can be effectively predicted using specific topological indices. This reinforces the practical significance of computational approaches in molecular property estimation and their potential application in drug discovery and development.

This paper is structured as follows: Section 2 provides an overview of preliminarily and graph model. The materials and methods employed in this study are explained in Section 3. Sections 4 and 5 present the analysis of linear and non-linear predictive modeling approaches, respectively. Finally, additional methodological frameworks and statistical analysis of the data are discussed in Section 6.

	\section{Preliminarily and Graph Model}
	A vertex and edge-weighted (VEW) molecular graph $\mathcal{G}$ is firstly defined in \cite{book,hand} as  
	$$\mathcal{G}=\mathcal{G}(V,E,Sym,Bo,Vw,Ew,w)$$ such that the vertex and edge set is $V= V(\mathcal{G})$ and $E =E(\mathcal{G})$. respectively;  Here a set of chemical symbols of the vertices $Sym=Sym(\mathcal{G})$, a set of topological bond orders ( takes the value 1 for single bonds, 2 for double bonds, 3 for triple bonds and 1.5 for aromatic bonds) of the edges $Bo = Bo(\mathcal{G})$, a vertex weight set $Vw(w) = V_w(w,\mathcal{G})$, and an edge weight set $Ew(w) = E_w(\mathcal{G})$. Here $w$ is the weighting scheme which is used to compute the $Vw(w)$ and $Ew(w)$. Generally, all schemes in a molecular graph are the properties of the atoms such as atomic number, atomic radius etc. \cite{hand}.
	
	\begin{equation} \label{vweighted}
		Vw(w)_{i}=1-\frac{w_{C}}{w_{i}}
	\end{equation}
	and 
	\begin{equation} \label{edgeweighted}
		Ew(w)_{ij}=\frac{w_{C}w_{C}}{Bo_{ij}w_{i}w_{j}}
	\end{equation}
	where $Vw(w)_{i}$ represents atom $i$ from a molecule; $Ew(w)_{ij}$ represents the bonds between atom $i$ and atom $j$ and $Bo_{ij}$ is the topological bonds order between $i$ and $j$, respectively \cite{book}.
	
	The adjacency matrix $A_w=A_w(\mathcal{G})$ of a vertex-edge-weighted molecular graph $G$ with $n$ vertices is the square $n\times n$ real symmetric matrix whose element $(A_w)_{uv}$  and $(D_w)_{uv}$ are defined in \cite{book} (pg. 173-175) as:
	
	\begin{equation} 
		(A_w)_{uv}=\begin{cases}
			V_w(w)_u   , & \text{if $u=v$}.\\
			E_w(w)_{uv}, & \text{if $uv\in E(G)$}\\
			0          , & \text{otherwise}\\
		\end{cases}
	\end{equation}
	and 
		\begin{equation} \label{distance}
		(D_w)_{uv}=\begin{cases}
			V_w(w)_u   , & \text{if $u=v$}.\\
			d_{w}(u,v), & \text{otherwise},\\
		\end{cases}
	\end{equation}
	respectively. Here $d_{w}(u,v)$ represents the distance between
	vertices $u$ and $v$ where  $w$ denotes the weighting scheme employed to calculate the parameters $V_w$ and $E_w$. In a VEW graph $G$, the length of a path $p_{ij}$ between vertices \( v_i \) and \( v_j \),
	\[
	l(p_{ij}, w) = l(p_{ij}, w, G),
	\] 
	is equal to the sum of the edge parameters \( Ew(w)_{ij} \) for all edges along the path.\\
	
	Topological indices are numerical descriptors derived from graphs, often calculated based on the elements of the graph. These indices frequently depend on properties such as vertex degrees or other structural characteristics, making them valuable tools for analyzing and interpreting molecular structures in various scientific fields. The application of topological indices in drug discovery has been well-documented in the literature, with numerous studies highlighting their effectiveness. \\
	
	Unlike the classical degree-distance-based topological indices, the topological definitions for VEW graphs are derived from equations in (\ref{vweighted}) and (\ref{edgeweighted}) as shown in the table below.
	
	\begin{table}[H]
		\caption{ VEW-based degree distance topological indices for $\mathcal{G}$}
		\label{3}
		\centering
		\resizebox{\textwidth}{!}{
			\begin{tabular}{l|l}
				\hline
				\textbf{TIs names} & \textbf{vew-based description} \\ \hline \hline
				Wiener & 	$W(G)=\sum_{u<v}(D_w)_{uv}$  \\ \hline
				Harary &$H(\mathcal{G})=\frac{1}{2}\sum_{u<v}\frac{1}{(D_w)_{uv}}$ \\ \hline 
				Balaban &$J(\mathcal{G})=\frac{m}{m-n+2}\sum_{uv\in E(\mathcal{G})}\frac{1}{\sqrt{(D_w)_u(D_w)_v}}$  \\ \hline
				Total Eccentric Index & $TEI(\mathcal{G})=\sum_{u\in V}\epsilon(u)$ \\ \hline
				Eccentric Connectivity Index & $ECI(\mathcal{G})=\sum_{u\in V}\epsilon(u)(A^{2}_w)_{uu}$ \\ \hline
				Degree Distance & $DD(\mathcal{G})=\sum_{uv\in E }[(A^{2}_w)_{uu}+(A^{2}_w)_{vv}](D_w)_{uv}$ \\ \hline
				Gutman Index & $G(\mathcal{G})=\sum_{uv\in E }[(A^{2}_w)_{uu}(A^{2}_w)_{vv}](D_w)_{uv}$ \\ \hline
				Reciprocal Distance Degree Index & $RDD(\mathcal{G})=\sum_{uv\in E }[(A^{2}_w)_{uu}+(A^{2}_w)_{vv}]/(D_w)_{uv}$ \\ \hline
		\end{tabular}}
	\end{table}
	In above table, notations of $(D_w)_u$ and $\epsilon(u)$ are the sum of all entries in the $u$th row of VEW distance matrix of graph $\mathcal{G}$ and  the maximum value of the $u$th row in the $D_w(G)$ matrix,respectively.\\
    
	For unweighted graphs, the classical definitions of distance and degree-distance-based topological indices, applications in chemistry and related between them can be found in  [\cite{Wiener}-\cite{RDD}] as listed in Table \ref{3}.

	\section{Material and Method }

	\subsection{Preparation of Data and Linear Regressions}
The molecular graphs of the 166 drugs listed in the following tabulars (Table \ref{ff} are obtained using their SMILES codes. Also, the properties of them have been taken from ChemSpyder database. Topological indices  is calculated by applying both vertex and edge weighting based on the atomic properties of the molecules, including atomic radius, atomic mass, density, ionization, electronegativity, and atomic number. For detailed information on the calculations, refer to \cite{sezer}.\\

\begin{table}

\begin{tabular}{|c|c|c|c|c|c|c|c|}
\hline
	No. & DRUGS & BP & MV & MR & FP & Polar & EV \\
	\hline
	1 & Chlorpromazine & 450.1 & 262.9 & 92.8 & 226 & 36.8 & 70.9 \\
	2 & Triphthasine & 506 & 328.8 & 108.2 & 259.8 & 42.9 & 77.6 \\
	3 & Thioridazine & 515.7 & 299.6 & 112.8 & 265.7 & 44.7 & 78.8 \\
	4 & Thiothixene & 599 & 349.4 & 126.5 & 316.1 & 50.1 & 89.2 \\
	5 & Haloperidol & 529 & 303.3 & 101 & 273.8 & 40 & 84.6 \\
	6 & Clozapine & 489.2 & 247.7 & 93.7 & 249.6 & 37.2 & 75.5 \\
	7 & Ziprasidone & 554.8 & 301.4 & 114.1 & 289.3 & 45.2 & 83.6 \\
	8 & Loxapine & 458.6 & 249.5 & 92.1 & 231.1 & 36.5 & 71.9 \\
	9 & Quetiapine & 556.5 & 301.1 & 110.2 & 290.4 & 43.7 & 88.2 \\
	10 & Qripiprazole & 646.2 & 355 & 120.3 & 344.6 & 47.7 & 95.3 \\
	11 & Risperidone & 572.4 & 296.8 & 111.7 & 300 & 44.3 & 85.8 \\
	12 & Olanzapine & 476 & 236 & 92.2 & 241.7 & 36.5 & 74 \\
	13 & Eliquis & 770.5 & 323.4 & 125.6 & 419.8 & 49.8 & 112.2 \\
	14 & Vericiguat & 535.9 & 260.8 & 104.8 & 277.9 & 41.5 & 81.2 \\
	15 & Dabigatran etexilate & 827.9 & 504 & 175.9 & 454.5 & 69.7 & 120.3 \\
	16 & Ivabradine & 626.9 & 408.7 & 132.2 & 332.9 & 52.4 & 92.8 \\
	17 & Dapagliflozin & 609 & 303.1 & 105.6 & 322.1 & 41.9 & 95.1 \\
	18 & Empagliflozin & 664.5 & 322.4 & 114.4 & 355.7 & 45.4 & 102.7 \\
	19 & Metoprolol & 398.6 & 258.7 & 77.1 & 194.9 & 30.6 & 68.5 \\
	20 & Sacubitril & 656.9 & 357.4 & 113.6 & 351.1 & 45 & 101.6 \\
	21 & Valsartan & 684.9 & 359.1 & 120.6 & 368 & 47.8 & 105.5 \\
	22 & Acarbose & 971.6 & 369.8 & 141.2 & 541.40 & 56 & 160.5 \\
	23 & Tolazamide & 484.5 & 237.90 & 82.5 & 246.80 & 32.7 & 79 \\
	24 & Miglitol & 453.7 & 142.10 & 48.5 & 284.30 & 19.2 & 82.3 \\
	25 & Prandin/Repaglinide & 672.9 & 397.90 & 130.1 & 360.80 & 51.6 & 103.8 \\
	\hline
\end{tabular}
\\
\caption{}
\label{ff}
\end{table}

\begin{table}
\begin{tabular}{l l r r r r r r}
\hline
	No. & DRUGS & BP & MV & MR & FP & Polar & EV \\
	\hline
	26 & Metformin & 172.5 & 100.80 & 33.4 & 58.10 & 13.2 & 40.9 \\
	27 & Linagliptin & 661.2 & 338.00 & 133.1 & 353.70 & 52.8 & 97.3 \\
	28 & Pioglitazone & 575.4 & 282.80 & 98.2 & 301.80 & 38.9 & 86.2 \\
	29 & Bromocriptine & 891.3 & 429.40 & 165.4 & 492.80 & 65.6 & 135.7 \\
	30 & Alogliptin & 519.2 & 252.90 & 93.3 & 267.80 & 37 & 79.2 \\
	31 & Chloroquine & 460.6 & 287.9 & 97.4 & 232.23 & 38.6 & 72.1 \\
	32 & Amodiaquine & 478 & 282.8 & 105.5 & 242.9 & 41.8 & 77 \\
	33 & Mefloquine & 415.7 & 273.4 & 83 & 205.2 & 32.9 & 70.5 \\
	34 & Piperoquine & 721.1 & 414.2 & 153.7 & 389.9 & 60.9 & 105.3 \\
	35 & Primaquine & 451.1 & 230.3 & 80.5 & 226.6 & 31.9 & 71 \\
	36 & Lumefantrine & 642.5 & 422.3 & 151 & 342.3 & 59.9 & 99.6 \\
	37 & Atovaquone & 535 & 271.8 & 99.5 & 277.3 & 39.5 & 85.4 \\
	38 & Pyrimethamine & 368.4 & 180.2 & 67.1 & 176.6 & 26.6 & 61.5 \\
	39 & Doxycycline (anhydrous) & 762.6 & 271.1 & 109 & 415 & 43.2 & 116.5 \\
	40 & Azathioprine & 685.7 & 145.4 & 68.9 & 368.5 & 27.3 & 96.9 \\
	41 & Hydroxychloroquine & 516.7 & 285.4 & 99 & 266.3 & 39.2 & 83 \\
	42 & Sulfasalazine & 689.3 & 267.7 & 102.4 & 370.7 & 40.6 & 106.1 \\
	43 & Filgotinib & -- & 281.1 & 114.3 & -- & 45.3 & -- \\
	44 & Leflunomide & 289.3 & 194.1 & 61 & 128.8 & 40.6 & 52.9 \\
	45 & Prednisolone & 570.6 & 274.7 & 95.5 & 313 & 37.9 & 98.3 \\
	46 & Methotrexate & -- & 295.7 & 119 & -- & 47.2 & -- \\
	47 & Baricitinib & 707.2 & 238.1 & 98.2 & 381.5 & 38.9 & 103.5 \\
	48 & Tofacitinib & 585.8 & 241 & 87.5 & 308.1 & 34.7 & 87.5 \\
	49 & Upadacitinib & -- & 243 & 91.6 & -- & 36.3 & -- \\
	50 & Fluticasone propionate & 568.3 & 377 & 121.1 & 297.5 & 48 & 98 \\
	51 & Clobetasone & 549 & 309.1 & 102.1 & 285.8 & 40.5 & 95.3 \\
	52 & Desonide & 580.1 & 320.1 & 109.3 & 196.9 & 43.3 & 99.6 \\
	53 & Clobetasol propionate & 569 & 364.1 & 117.8 & 297.9 & 46.7 & 98.1 \\
	54 & Azathioprine & 685.7 & 145.4 & 68.9 & 368.5 & 27.3 & 96.9 \\
	55 & Monobenzone & 359.1 & 172.6 & 59.3 & 213.4 & 23.5 & 62.8 \\
	56 & Betamethasone valerate & 598.9 & 382.4 & 123.7 & 316 & 49 & 102.3 \\
	57 & Psoralen & 362.6 & 134 & 49.9 & 173.1 & 19.8 & 60.9 \\
	58 & Hydrocortisone valerate & 595.1 & 367.6 & 119 & 195 & 47.2 & 101.8 \\
	59 & Fluticasone & 553.2 & 323.2 & 106.9 & 288.4 & 42.4 & 95.9 \\
	60 & Cidofovir & 609.5 & 158.6 & 58.3 & 322.4 & 23.1 & 103.8 \\
	61 & Foscarnet & 490.7 & 58.8 & 18.2 & 250.6 & 7.2 & 82.9 \\
	62 & Maribavir & 611 & 224 & 86.9 & 323.3 & 34.4 & 95.4 \\
	63 & Valganciclovir & 629.1 & 222.5 & 83.9 & 334.3 & 33.3 & 97.8 \\
	64 & Dacomitinib & 665.7 & 349.5 & 129.5 & 356.4 & 51.3 & 97.9 \\
	65 & Selpercatinib & -- & 383.9 & 147.5 & -- & 58.5 & -- \\
	66 & Tepotinib & 626.5 & 391.6 & 144.5 & 332.7 & 57.3 & 92.7 \\
	67 & Sotorasib & 730.5 & 411.9 & 150.5 & 395.6 & 59.6 & 110.4 \\
	68 & Etoposide & 798.1 & 378.5 & 140.1 & 263.6 & 55.5 & 121.7 \\
	69 & Alectinib & 722.5 & 374.7 & 140.4 & 390.7 & 55.7 & 105.5 \\
	70 & Paclitaxel & 957.1 & 610.6 & 219.3 & 532.6 & 86.9 & 146 \\
	\hline
\end{tabular} \\
\end{table}
\begin{table}
\begin{tabular}{l l r r r r r r}
	\hline
	No. & DRUGS & BP & MV & MR & FP & Polar & EV \\
	\hline
71 & Dabrafenib & 653.7 & 359.9 & 127.4 & 349.2 & 50.5 & 96.3 \\
72 & Entrectinib & 717.5 & 418.1 & 156.6 & 387.7 & 62.1 & 104.8 \\
73 & Crizotinib & 599.2 & 305.2 & 114.4 & 316.2 & 45.4 & 89.2 \\
74 & Ceritinib & 720.7 & 446 & 151.5 & 389.6 & 60.1 & 105.3 \\
75 & Lorlatinib & 675 & 285 & 108.5 & 362.1 & 43 & 99.1 \\
76 & Afatinib & 676.9 & 352 & 131.2 & 363.2 & 52 & 99.4 \\
77 & Pralsetinib & 799.1 & 381 & 144.5 & 437.1 & 57.3 & 116.2 \\
78 & Brigatinib & 781.8 & 443.6 & 160.1 & 426.6 & 63.5 & 113.8 \\
79 & Erlotinib & 553.6 & 315.4 & 110.1 & 288.6 & 43.6 & 83.4 \\
80 & Adagrasib & 860.2 & 466.2 & 163.4 & 474 & 64.8 & 125 \\
81 & Gefitinib & 586.8 & 337.8 & 118.8 & 308.7 & 47.1 & 87.6 \\
82 & Vinorelbine & -- & 569.7 & 214.2 & -- & 84.9 & -- \\
83 & Gemcitabine & 482.7 & 142.3 & 52.1 & 245.7 & 20.6 & 86.2 \\
84 & Docetaxel & 900.5 & 585.7 & 205.2 & 498.4 & 81.4 & 137.1 \\
85 & Pemetrexed & -- & 268.1 & 106.3 & -- & 42.1 & -- \\
86 & Gefitinib & 586.8 & 337.8 & 118.8 & 308.7 & 47.1 & 87.6 \\
87 & Erlotinib & 553.6 & 315.4 & 110.1 & 288.6 & 43.6 & 83.4 \\
88 & Canertinib & 691 & 358.5 & 130.7 & 371.7 & 51.8 & 101.3 \\
89 & Afatinib & 676.9 & 352 & 131.2 & 363.2 & 52 & 99.4 \\
90 & Vandetanib & 538.2 & 338 & 120 & 279.3 & 47.6 & 81.5 \\
91 & Ispinesib & 708 & 425.2 & 149.2 & 382 & 59.2 & 103.6 \\
92 & Tacrine & 353.8 & 157.8 & 59.8 & 167.8 & 23.7 & 59.9 \\
93 & Donepezil & 527.9 & 332.5 & 110.4 & 273.1 & 43.8 & 80.3 \\
94 & Rivastigmine & 316.2 & 241.2 & 73.1 & 145 & 29 & 55.8 \\
95 & Galantamine & 439.3 & 223.9 & 80.3 & 219.5 & 31.8 & 73.4 \\
96 & Huperzine A & 505 & 201.8 & 71.5 & 259.2 & 28.3 & 77.5 \\
97 & Amikacin & 981.8 & 363.9 & 134.9 & 547.6 & 53.5 & 162.2 \\
98 & Bedaquiline & 702.7 & 420.1 & 156.2 & 378.8 & 61.9 & 108 \\
99 & Clofazimine & 566.9 & 366.1 & 136.2 & 296.7 & 54 & 85.1 \\
100 & Delamanid & 653.7 & 368 & 127.7 & 349.1 & 50.6 & 96.3 \\
101 & Ethambutol & 345.3 & 207 & 58.6 & 113.7 & 23.2 & 68.3 \\
102 & Ethionamide & 247.9 & 142 & 49 & 103.7 & 19.4 & 46.5 \\
103 & Imipenem-cilastatin & 530.2 & 183.9 & 72.7 & 274.5 & 28.8 & 92.7 \\
104 & Levofloxacin & 571.5 & 244 & 91.1 & 299.4 & 36.1 & 90.1 \\
105 & Linezolid & 585.5 & 259 & 83 & 307.9 & 32.9 & 87.5 \\
106 & Moxifloxacin & 636.4 & 285 & 101.8 & 338.7 & 40.4 & 98.8 \\
107 & p-Aminosalicylic acid & 380.8 & 102.7 & 39.3 & 184.1 & 15.6 & 66.3 \\
108 & Pyrazinamide & 273.3 & 87.7 & 31.9 & 119.1 & 12.6 & 54.1 \\
109 & Rifampin & 1004.4 & 611.7 & 213.1 & 561.3 & 84.5 & 153.5 \\
110 & Terizidone & -- & 198.9 & 76.1 & -- & 30.2 & -- \\
111 & Cefuroxime & 731.7 & 241.0 & 96.7 & 396.3 & 38.3 & 112.1 \\	
\hline
\end{tabular}
\end{table}
\begin{table}
\begin{tabular}{l l r r r r r r}
	\hline
	No. & DRUGS & BP & MV & MR & FP & Polar & EV \\
	\hline
112 & Amoxicillin & 743.2 & 236.2 & 91.5 & 403.3 & 36.2 & 113.7 \\
113 & Ofloxacin & 571.5 & 244.0 & 91.1 & 299.4 & 36.1 & 90.1 \\
114 & Cortisol & 566.5 & 281.4 & 95.6 & 310.4 & 37.9 & 97.9 \\
115 & Moxifloxacin & 636.4 & 285.0 & 101.8 & 338.7 & 40.4 & 98.8 \\
116 & Flurandrenolide & 578.8 & 322.3 & 109.4 & 303.8 & 43.4 & 99.4 \\
117 & Dapsone & 511.7 & 182.4 & 67.5 & 263.2 & 26.8 & 78.3 \\
118 & Propranolol & 434.9 & 237.2 & 79.0 & 216.8 & 31.3 & 72.8 \\
119 & Metoprolol & 398.6 & 258.7 & 77.1 & 194.9 & 30.6 & 68.5 \\
120 & DL-Atenolol & 508.0 & 236.7 & 74.3 & 261.1 & 29.4 & 81.9 \\
121 & Carvedilol & 655.2 & 325.1 & 119.6 & 350.1 & 47.4 & 101.4 \\
122 & Azacitidine & 534.21 & 117.1 & 51.1 & 277 & 20.3 & 93.2 \\
123 & Busulfan & 464 & 182.4 & 50.9 & 234.4 & 20.2 & 69.8 \\
124 & Mercaptopurine & 491 & 94.2 & 41 & 250.5 & 16.2 & 72.8 \\
125 & Tioguanine & 460.7 & 80.2 & 41.9 & 232.4 & 16.6 & 72.1 \\
126 & Nelarabine & 721 & 149.9 & 65.8 & 389.9 & 26.1 & 110.6 \\
127 & Cytarabine & 547.7 & 128.4 & 52 & 283.8 & 20.9 & 94.8 \\
128 & Clofarabine & 550 & 143.1 & 63.6 & 316.4 & 25.2 & 93.9 \\
129 & Bosutinib & 649.7 & 388.3 & 141.9 & 346.7 & 56.3 & 95.8 \\
130 & Dasatinib & 133.08 & 366.4 & 132 & -- & 52.3 & -- \\
131 & Melphalan & 473 & 231.2 & 78.8 & 239.9 & 31.2 & 77.6 \\
132 & Dexamethasone & 568.2 & 296.2 & 100.2 & 297.5 & 39.7 & 98 \\
133 & Doxorubicine & 810.3 & 336.6 & 131.5 & 443.8 & 52.1 & 123.5 \\
134 & Amathaspiramide E & 572.7 & 233.9 & 89.4 & 300.2 & 35.4 & 90.3 \\
135 & Aminopterin & 782.27 & 277.2 & 114.3 & -- & 45.3 & -- \\
136 & Aspidostomide & 798.8 & 262 & 116 & 436.9 & 46 & 116.2 \\
137 & Carmustine & 309.6 & 146.4 & 46.6 & 141 & 18.5 & 63.8 \\
138 & Caulibugulone E & 373 & 139.1 & 52.2 & 179.4 & 20.7 & 62 \\
139 & ConvolutamideA & 629.9 & 396 & 130.1 & 334.7 & 51.6 & 97.9 \\
140 & ConvolutamineF & 387.7 & 220.1 & 73.8 & 188.3 & 29.2 & 63.7 \\
141 & Convolutamydine A & 504.9 & 190 & 68.2 & 259.2 & 27 & 81.6 \\
142 & Daunorubicin & 770 & 339.4 & 130 & 419.5 & 51.5 & 117.6 \\
143 & Deguelin & 560.1 & 314.2 & 105.1 & 244.7 & 41.7 & 84.3 \\
144 & Melatonin & 512.8 & 197.6 & 67.6 & 264 & 26.8 & 78.4 \\
145 & Minocycline & 803.3 & 294.6 & 116 & 439.6 & 46 & 122.5 \\
146 & Perfragilin A & 431.5 & 167.8 & 63.6 & 214.8 & 25.2 & 68.7 \\
147 & Pterocellin B & 521.6 & 228.3 & 87.4 & 269.2 & 34.7 & 79.5 \\
148 & Raloxifene & 728.2 & 367.3 & 136.6 & 394.2 & 54.1 & 110.1 \\
149 & Abemaciclib & 689.3 & 382.3 & 140.4 & 370.7 & 55.6 & 101 \\
150 & Paclitaxel & 957.1 & 610.6 & 219.3 & 532.6 & 86.9 & 146 \\
151 & Anastrozole & 469.7 & 270.3 & 90 & 237.9 & 35.7 & 73.2 \\
152 & Capecitabine & 517 & 240.5 & 82.3 & -- & 32.6 & -- \\
153 & Cyclophosphamide & 336.1 & 195.7 & 58.1 & 157.1 & 23 & 57.9 \\
154 & Everolimus & 998.7 & 811.2 & 257.7 & 557.8 & 102.2 & 165.1 \\
155 & Exemestane & 453.7 & 260.6 & 85.8 & 169 & 34 & 71.3 \\
156 & Fulvestrant & 674.8 & 505.1 & 154 & 361.9 & 61.1 & 104.1 \\
157 & Ixabepilone & 697.8 & 451.6 & 140.1 & 375.8 & 55.5 & 107.3 \\
158 & Letrozole & 563.5 & 234.5 & 87.1 & 294.6 & 34.5 & 84.7 \\
159 & Megestrol Acetate & 507.1 & 333.4 & 106.4 & 218.5 & 42.2 & 77.7 \\
160 & Methotrexate & -- & 295.7 & 119 & -- & 47.2 & -- \\
161 & cis-Tamoxifen & 482.3 & 118.9 & 118.9 & 140 & 47.1 & 74.7 \\
162 & Thiotepa & 270.2 & 125.8 & 49.1 & 117.2 & 19.5 & 50.8 \\
163 & Glasdegib & 633.4 & 281 & 106.9 & 336.9 & 42.4 & 93.6 \\
164 & Palbociclib & 711.5 & 340.7 & 123.9 & 384.1 & 49.1 & 104 \\
165 & Gilteritinib & 696.9 & 444.9 & 157.8 & 375.3 & 62.5 & 102.1 \\
166 & Ivosidenib & 854.3 & 383.6 & 140.1 & 470.4 & 55.5 & 124.1 \\
\hline
\end{tabular}
\end{table}

In the following sections, we introduce both linear and nonlinear modeling approaches to establish the relationship between six physical properties—Boiling Point (BP), Molar Volume (MV), Molar Refraction (MR), Flash Point (FP), Polarizability (Polar), and Enthalpy of Vaporization (EV)—and eight topological indices mentioned in Table \ref{3}. For linear models, we employ Linear Regression, Lasso Regression, and Ridge Regression, while for nonlinear approaches, we utilize Random Forest, XGBoost, and Neural Networks. For more details on machine learning methods, refer to \cite{bishop}. These methodologies enable a comprehensive evaluation of the predictive capabilities of topological indices in capturing physicochemical properties. 

\subsection{Linear and Non-Linear Regression Methods}  
In this paper, Linear regression and its variants—Lasso and Ridge regression—are extensively employed in quantitative structure–property relationship (QSPR) analyses, aiming to predict molecular physical properties from their topological indices.  \\

\textbf{Linear Regression:}  
Linear regression establishes a direct mathematical relationship between a dependent variable (physical property) and multiple independent variables (topological indices). Given a dataset with \( n \) molecular descriptors, the regression model is expressed as:  

$$
Y = \beta_0 + \beta_1 X_1 + \beta_2 X_2 + ... + \beta_n X_n + \epsilon
$$  
where $Y$ represents the predicted physical property, $X_i$ are the topological indices, $\beta_i$ are the regression coefficients, and $\epsilon$ is the error term. While simple and interpretable, linear regression can suffer from overfitting when multicollinearity exists among descriptors.  

\textbf{Lasso Regression:}  
Lasso (Least Absolute Shrinkage and Selection Operator) regression is a modified form of linear regression that incorporates an \( L_1 \) penalty term in the cost function:  

$$
\min_{\beta} \sum_{i=1}^{m} (Y_i - \sum_{j=1}^{n} \beta_j X_{ij})^2 + \lambda \sum_{j=1}^{n} |\beta_j|.  $$

The inclusion of the $L_1$-norm forces some coefficients to become exactly zero, effectively performing variable selection. This is particularly beneficial in cases where only a subset of the topological indices contributes significantly to the prediction of molecular properties, reducing model complexity and improving interpretability.  

\textbf{Ridge Regression:}  
Ridge regression addresses multicollinearity by adding an $L_2$-norm penalty to the linear regression cost function:  

$$
\min_{\beta} \sum_{i=1}^{m} (Y_i - \sum_{j=1}^{n} \beta_j X_{ij})^2 + \lambda \sum_{j=1}^{n} \beta_j^2.  $$

Unlike Lasso, Ridge regression does not set coefficients to zero but shrinks them toward smaller values, preventing overfitting while retaining all predictor variables. This is particularly useful when all topological indices provide complementary information on molecular properties. \\

We employed Random Forest, XGBoost, and Neural Networks as nonlinear modeling approaches to predict the physical properties of molecules based on distance-degree-based topological indices. In the following we explain them briefly. \\

\textbf{Random Forest:}
Random Forest (RF) is an ensemble learning method that constructs multiple decision trees during training and aggregates their predictions to enhance accuracy and reduce overfitting. Each tree is built using a random subset of the training data and a random selection of predictor variables, improving model generalization. The prediction for a given input is determined by averaging (for regression) or majority voting (for classification) across all trees. The RF model can be expressed as:  

$$
Y = \frac{1}{T} \sum_{t=1}^{T} f_t(X),
$$
where $T$ represents the number of trees, $f_t(X)$ is the prediction from the $t$-th tree, and $Y$ is the final predicted physical property. RF is particularly advantageous for handling high-dimensional data with complex interactions between features, such as topological indices, while being robust to noise and multicollinearity. \\

\textbf{XGBoost:}  
Extreme Gradient Boosting (XGBoost) is an optimized gradient boosting algorithm designed to improve prediction accuracy and computational efficiency. Unlike Random Forest, which trains trees independently, XGBoost builds trees sequentially, with each new tree correcting errors from the previous iterations. The model minimizes a regularized objective function:  

$$
\mathcal{L} = \sum_{i=1}^{m} l(Y_i, \hat{Y}_i) + \sum_{t=1}^{T} \Omega(f_t),
$$

where $l(Y_i, \hat{Y}_i)$ is the loss function, measuring the difference between observed and predicted values, and $\Omega(f_t)$ is the regularization term that prevents overfitting. XGBoost effectively captures nonlinear relationships between molecular descriptors and physicochemical properties, making it a powerful tool in quantitative structure-property relationship (QSPR) modeling.\\  

\textbf{Neural Networks}  
Artificial Neural Networks (ANNs) are computational models inspired by the structure of biological neurons. They consist of multiple interconnected layers of neurons that process and transform input features through weighted connections and activation functions. A typical feedforward neural network with one hidden layer can be mathematically represented as:  

$$
Y = \sigma \left( W_2 \cdot \sigma(W_1 X + b_1) + b_2 \right),
$$

where $X$ represents the input topological indices, $W_1$ and $W_2$ are weight matrices, $b_1$ and $b_2$ are bias terms, and $\sigma$ is the activation function (e.g., ReLU or sigmoid). The network is trained using backpropagation and gradient-based optimization to minimize the prediction error. ANNs excel in capturing highly nonlinear relationships and intricate feature interactions, making them suitable for modeling complex molecular property predictions.  

\subsection{Type of Metric Validation}
In regression analysis, the $R^2$ (coefficient of determination), RMSE (Root Mean Squared Error), and MAE (Mean Absolute Error) are standard metrics used to evaluate model performance.  
\begin{itemize}
    \item $R^2$ Score: Measures the proportion of variance in the dependent variable that is explained by the independent variables. An $R^2$ value closer to 1 indicates a better fit, meaning the model effectively captures the relationship between features and the target variable.  
\item RMSE (Root Mean Squared Error): Represents the square root of the average squared differences between predicted and actual values. It penalizes large errors more than MAE and provides an overall measure of model accuracy, with lower values indicating better performance.  

\item MAE (Mean Absolute Error): Calculates the average absolute difference between predicted and actual values, offering an interpretable measure of prediction accuracy. Unlike RMSE, it treats all errors equally without emphasizing larger deviations.  

\end{itemize}

Together, these metrics provide a comprehensive assessment of model reliability, with $R^2 $ indicating explanatory power and RMSE/MAE quantifying prediction errors.

\section{Linear Regression Prediction}  
For the dataset of 166 drugs analyzed in this study, these regression techniques allow the identification of the most relevant topological indices to predict six different physical properties. Ridge regression provides stability in the presence of correlated indices, while Lasso facilitates the selection of the most informative descriptors, enhancing model interpretability. The choice between these methods depends on the complexity and redundancy of the molecular descriptors, balancing predictive performance with scientific insight. 

\begin{figure}
    \centering
    \begin{minipage}{0.8\textwidth}
        \centering
    \includegraphics[width=\textwidth]{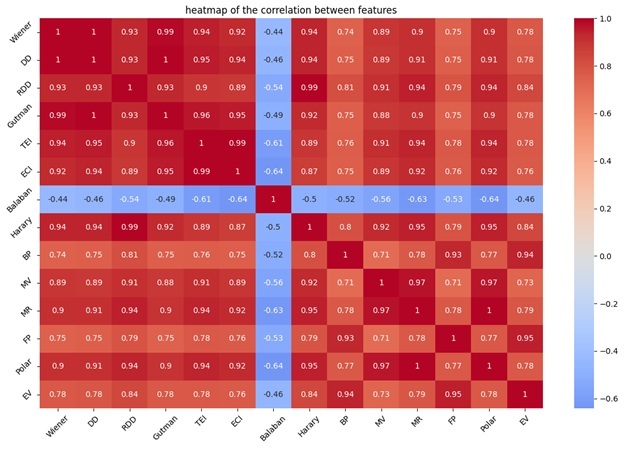}
  \end{minipage}
    \caption{Heatmap of the correlation between features}
    \label{corrolation}
    
\end{figure}

In order to address missing data, the K-nearest neighbors (KNN) algorithm was used for imputation. Furthermore, the data set was normalized and the five topological indices exhibiting the strongest correlation were selected for regression analysis (see, Fig. \ref{corrolation}). Regularization techniques were applied to optimize model parameters, and cross-validation was performed to mitigate overfitting. 

\subsection{Comprehensive Analysis of Linear Regression Performance Based on Atomic Number and Atomic Mass}  

The predictive performance of regression models in estimating six physicochemical properties Boiling Point (BP), Molar Volume (MV), Molar Refraction (MR), Flash Point (FP), Polarizability (Polar), and Enthalpy of Vaporization (EV) was evaluated using topological indices derived from atomic number and atomic mass. The results, as summarized in Tables \ref{analAN} and \ref{analAM}, provide insight into the effectiveness of various regression models and the contribution of different topological indices to property prediction.  
\subsubsection{Model Selection and Predictive Accuracy}

    Across both datasets, Ridge regression consistently demonstrated superior performance for five out of the six target properties (BP, MR, FP, Polar, and EV), highlighting its robustness in handling collinear molecular descriptors. For MV, LASSO regression was identified as the best-performing model, emphasizing the role of feature selection in improving predictive performance for this property.  

     The highest $R^2$ values were observed for MR and Polar in both atomic number and atomic mass-based analyses ($R^2 \approx 0.95$), indicating that these properties exhibit strong correlations with topological indices. MV also demonstrated a high  $R^2$ score ($\sim 0.88$), reflecting reliable predictability. In contrast, FP showed the lowest \( R^2 \) values (~0.656), suggesting a more complex underlying relationship that is less effectively captured by the regression models.  
     
   Error Metrics The lowest Root Mean Squared Error (RMSE) and Mean Absolute Error (MAE) were obtained for MR and Polar, reinforcing the strong predictive capacity for these properties. Conversely, FP exhibited the highest RMSE and MAE, consistent with its lower \( R^2 \), indicating greater variability and prediction difficulty.

   While both atomic number and atomic mass-based topological indices produced similar predictive trends, slight variations in correlation values and model performance were observed. Atomic mass-based indices yielded marginally better predictive accuracy for MR and Polar ($ R^2 \approx 0.9501$), whereas atomic number-based indices produced slightly improved predictions for BP and EV. These variations suggest that different molecular representations may enhance predictive capabilities for specific physicochemical properties.

\subsubsection{Influence of Topological Indices on Property Prediction}  

The correlation analysis of topological indices with physicochemical properties provides valuable insights into their predictive significance:  
\begin{itemize}
    \item  Reciprocal Distance Degree Index exhibited the highest correlation with MR ($\sim 0.951$) and Polar ($\sim 0.950$), suggesting that this index effectively captures molecular features influencing these properties.
    \item Total Eccentric Index and Eccentric Connectivity Index demonstrated strong correlations with MV, MR, and Polar, further confirming their predictive importance.
    \item The Harary index showed significant correlations with BP (0.818) and EV (0.852), suggesting its potential role in modeling these thermal properties.
    \item Wiener index exhibited a moderate correlation with EV (0.779) in the atomic mass-based analysis, indicating partial influence on this property but lower predictive strength compared to other indices.
\end{itemize}

These findings emphasize the importance of selecting appropriate regression techniques and molecular descriptors for accurate Quantitative Structure-Property Relationship (QSPR) modeling. The strong correlations observed between topological indices and physicochemical properties highlight the potential of using graph-based molecular descriptors in cheminformatics applications.

\begin{table}[h]
	\centering
	\caption{ Comprehensive Analysis of Results (Atomic Number) }
	\small
	\begin{tabular}{@{}l*{6}{>{\centering\arraybackslash}p{2cm}}@{}}
		\toprule
		\textbf{Metric/TI} & \textbf{BP} & \textbf{MV} & \textbf{MR} & \textbf{FP} & \textbf{Polar} & \textbf{EV} \\
		\midrule
		\rowcolor{lightgray}
		Best Model & Ridge & LASSO & Ridge & Ridge & Ridge & Ridge \\
		$R^2$ Score & 0.7377 & 0.8774 & 0.9495 & 0.6562 & 0.9495 & 0.7420 \\
		RMSE & 0.4920 & 0.3476 & 0.2201 & 0.5738 & 0.2210 & 0.4981 \\
		MAE & 0.3905 & 0.2706 & 0.1641 & 0.4399 & 0.1644 & 0.3922 \\
		\midrule
		
		\rowcolor{lightgray}
		\multicolumn{6}{l}{\textbf{$TIs$ (Correlations)}} \\
		Harary & 0.8183 & - & - & 0.8101 & - & 0.8524 \\
		RDD & 0.8054 & 0.9161 & 0.9514 & 0.7903 & 0.9499 & 0.8356 \\
		TEI & 0.7655 & 0.9259 & 0.9468 & 0.7694 & 0.9458 & 0.7791 \\
		ECI & - & 0.9273 & 0.9514 & - & 0.9502 & - \\
		\bottomrule
        \label{analAN}
	\end{tabular}
	\end{table}

\begin{table}[h]
	\centering
	\caption{ Comprehensive Analysis of Results (Atomic Mass) }
	\small
	\begin{tabular}{@{}l*{6}{>{\centering\arraybackslash}p{2cm}}@{}}
		\toprule
		\textbf{Metric/TI} & \textbf{BP} & \textbf{MV} & \textbf{MR} & \textbf{FP} & \textbf{Polar} & \textbf{EV} \\
		\midrule
		\rowcolor{lightgray}
		Best Model & Ridge & LASSO & Ridge & Ridge & Ridge & Ridge \\
		$R^2$ Score & 0.7373 & 0.8773 & 0.9501 & 0.6563& 0.9501 & 0.7414 \\
		RMSE     & 0.4925 & 0.3478 & 0.2189 & 0.5741 & 0.2198& 0.4986 \\
		MAE      & 0.3910 & 0.2707 & 0.1633 & 0.4402 & 0.1636 & 0.3924 \\
		\midrule
		
		\rowcolor{lightgray}
		\multicolumn{6}{l}{\textbf{$TIs$ (Correlations)}} \\
		Harary & 0.8183 & -     & -      & 0.8102 & -      & 0.8524 \\
		RDD & 0.8057   & 0.9162 & 0.9516 & 0.7907 & 0.9501& 0.8358 \\
		TEI & 0.7656   & 0.9259& 0.9468 & 0.7694 & 0.9458 & - \\
		ECI & -        & 0.9273& 0.9515  & -      & 0.9503 & - \\
		Wiener & -        & -& -  & -      & - & 0.7790 \\
		\bottomrule
        \label{analAM}
	\end{tabular}
	
	\vspace{2mm}
\footnotesize
	{Note: All correlation coefficients are in the $[0,1]$ range. The dash (-) denotes absence of significant topological indices for that parameter.}
	\end{table}

In addition to evaluating model performance using  $R^2$, RMSE, and MAE scores, a detailed analysis of regression coefficients provides insights into the contribution of individual topological indices to the prediction of physicochemical properties. Table \ref{coeff} presents the regression coefficients for selected features, highlighting the most influential descriptors in each model. The significance of these coefficients varies across different properties, emphasizing the varying degrees of linear correlation between molecular structure and physical attributes.  

\begin{figure}
    \centering
    \begin{minipage}{0.5\textwidth}
        \centering
        \includegraphics[width=\textwidth]{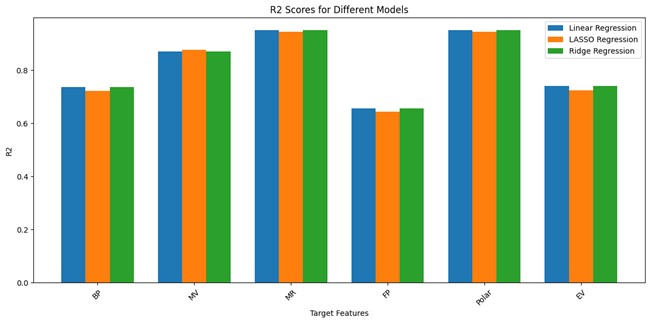}
    \end{minipage}
    \begin{minipage}{0.5\textwidth}
        \centering
        \includegraphics[width=\textwidth]{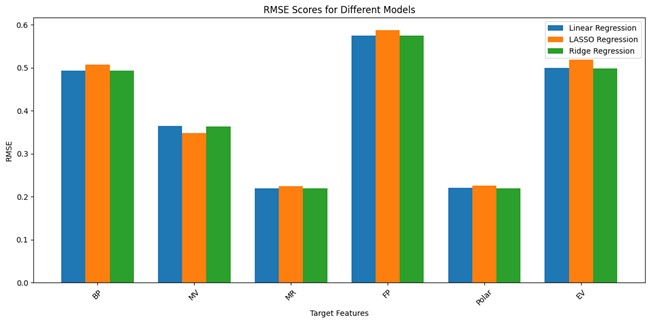}
        \end{minipage}
      \begin{minipage}{0.5\textwidth}
        \centering
        \includegraphics[width=\textwidth]{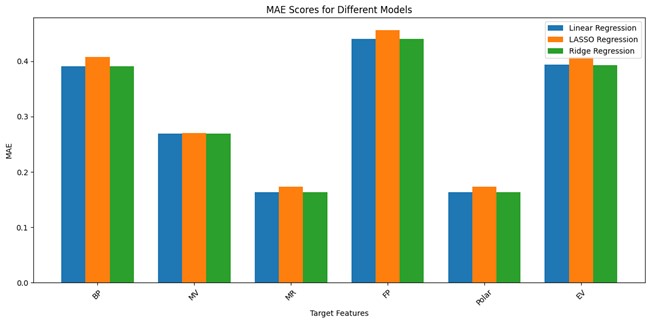}  
    \end{minipage}
    \caption{$R^2$, $RMSE$ and $MAE$ scores for different models based on Atomic Mass}
    \label{AMreg}
\end{figure}

\begin{figure}
    \centering
    \begin{minipage}{0.5\textwidth}
       \centering
        \includegraphics[width=\textwidth]{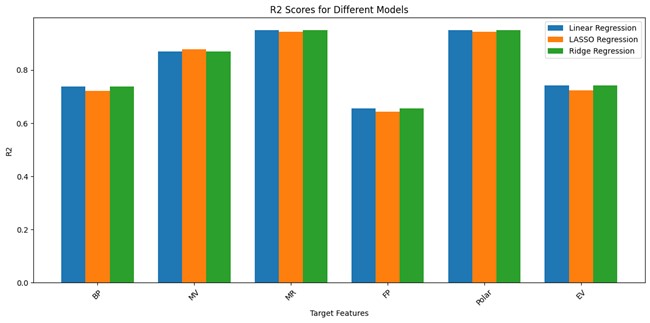}
    \end{minipage}
    \begin{minipage}{0.5\textwidth}
       \centering
       \includegraphics[width=\textwidth]{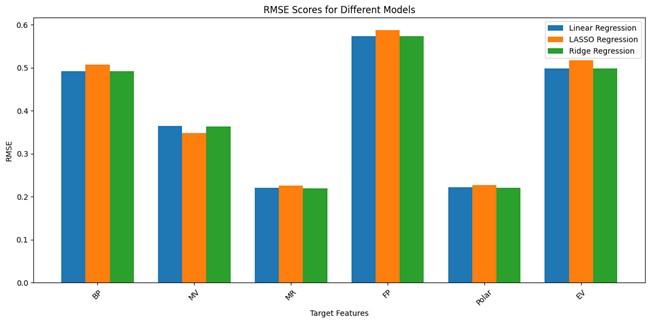}
       \end{minipage}
      \begin{minipage}{0.5\textwidth}
        \centering
        \includegraphics[width=\textwidth]{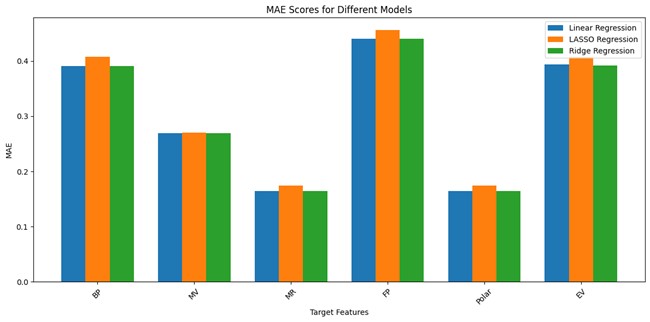}  
    \end{minipage}
        \caption{$R^2$, $RMSE$ and $MAE$ scores for different models based on Atomic Number}
        \label{ARreg}
\end{figure}

Furthermore, Figures \ref{AMreg} and \ref{ARreg} illustrate the comparative performance of different regression models based on atomic mass and atomic number, respectively. The $R^2$ values indicate the extent to which each model explains the variance in the target properties, while RMSE and MAE errors provide a measure of predictive accuracy and error magnitude. A comparative assessment of these results enables a deeper understanding of the strengths and limitations of both linear and non-linear models in cheminformatics applications.

\begin{table}[H]
\centering
\caption{Regression Coefficients for selected features}
\label{coeff}
\scriptsize

\begin{minipage}[t]{0.45\textwidth}
\centering
\begin{tabular}{@{}l*{3}{S}@{}}
\toprule
\multirow{2}{*}{Parameter/Feature} & \multicolumn{3}{c}{atomic number scheme} \\
\cmidrule(lr){2-4}
 & {Linear} & {LASSO} & {Ridge} \\
\midrule

\rowcolor{lightgray}
\multicolumn{4}{@{}l}{\textbf{BP}} \\
Wiener & -0.7053 & -0.3399 & -0.7023 \\
RDD & 0.1060 & \zero & 0.0934 \\
TEI & 1.6706 & 0.3723 & 1.6379 \\
ECI & -1.1059 & \zero & -1.0744 \\
Harary & 0.8495 & 0.7787 & 0.8606 \\
\addlinespace[2mm]


\bottomrule
\end{tabular}
\end{minipage}
\hspace{0.02\textwidth} 
\begin{minipage}[t]{0.45\textwidth}
\centering
\begin{tabular}{@{}l*{3}{S}@{}}
\toprule
\multirow{2}{*}{Parameter/Feature} & \multicolumn{3}{c}{atomic mass scheme} \\
\cmidrule(lr){2-4}
 & {Linear} & {LASSO} & {Ridge} \\
\midrule

\rowcolor{lightgray}
\multicolumn{4}{@{}l}{\textbf{BP}} \\
Wiener & -0.6991 & -0.3383 & -0.6962 \\
RDD & 0.1232 & \zero & 0.1105 \\
TEI & 1.6634 & 0.3733 & 1.6307 \\
ECI & -1.1014 & \zero & -1.0699 \\
Harary & 0.8293 & 0.7765 & 0.8404 \\
\addlinespace[2mm]


\bottomrule
\end{tabular}
\end{minipage}

\vspace{2mm}
\begin{minipage}[t]{0.45\textwidth}
\centering
\begin{tabular}{@{}l*{3}{S}@{}}
\toprule
\multirow{2}{*}{Parameter/Feature} & \multicolumn{3}{c}{atomic number scheme} \\
\cmidrule(lr){2-4}
 & {Linear} & {LASSO} & {Ridge} \\
\midrule

\rowcolor{lightgray}
\multicolumn{4}{@{}l}{\textbf{MV}} \\
DD & -0.2040 & \zero & -0.2037 \\
RDD & 0.4576 & 0.4264 & 0.4552 \\
TEI & 0.3577 & 0.2540 &0.3545 \\
ECI & 0.2953 & 0.2853& 0.2983 \\
Harary &0.0723 & \zero & 0.0746 \\
\addlinespace[2mm]


\bottomrule
\end{tabular}
\end{minipage}
\hspace{0.02\textwidth} 
\begin{minipage}[t]{0.45\textwidth}
\centering
\begin{tabular}{@{}l*{3}{S}@{}}
\toprule
\multirow{2}{*}{Parameter/Feature} & \multicolumn{3}{c}{atomic mass scheme} \\
\cmidrule(lr){2-4}
 & {Linear} & {LASSO} & {Ridge} \\
\midrule

\rowcolor{lightgray}
\multicolumn{4}{@{}l}{\textbf{MV}} \\
DD & -0.2074 & \zero & -0.2071 \\
RDD & 0.4520& 0.4271 & 0.4496 \\
TEI & 0.3558 &0.2594 & 0.3528 \\
ECI & 0.2984 & 0.2791 & 0.3013 \\
Harary & 0.0803 & \zero & 0.0825 \\
\addlinespace[2mm]


\bottomrule
\end{tabular}
\end{minipage}

\vspace{2mm}
\begin{minipage}[t]{0.45\textwidth}
\centering
\begin{tabular}{@{}l*{3}{S}@{}}
\toprule
\multirow{2}{*}{Parameter/Feature} & \multicolumn{3}{c}{atomic number scheme} \\
\cmidrule(lr){2-4}
 & {Linear} & {LASSO} & {Ridge} \\
\midrule

\rowcolor{lightgray}
\multicolumn{4}{@{}l}{\textbf{MR}} \\
RDD & 0.4319 & 0.5331 & 0.4389 \\
Gutman & -0.4033 & -0.0908 & -0.4015 \\
TEI & -0.2333 & \zero & -0.2184 \\
ECI & 0.9425 & 0.5399 & 0.9264 \\
Harary & 0.2617 & \zero & 0.2538 \\
\addlinespace[2mm]


\bottomrule
\end{tabular}
\end{minipage}
\hspace{0.02\textwidth} 
\begin{minipage}[t]{0.45\textwidth}
\centering
\begin{tabular}{@{}l*{3}{S}@{}}
\toprule
\multirow{2}{*}{Parameter/Feature} & \multicolumn{3}{c}{atomic mass scheme} \\
\cmidrule(lr){2-4}
 & {Linear} & {LASSO} & {Ridge} \\
\midrule

\rowcolor{lightgray}
\multicolumn{4}{@{}l}{\textbf{MR}} \\
RDD & 0.4317 & 0.5349 & 0.4388 \\
Gutman & -0.4048 & -0.0934 & -0.4030 \\
TEI & -0.2443 & \zero & -0.2290 \\
ECI & 0.9541 & 0.5409 & 0.9377 \\
Harary & 0.2630 & \zero & 0.2550 \\
\addlinespace[2mm]


\bottomrule
\end{tabular}
\end{minipage}

\vspace{2mm}
\begin{minipage}[t]{0.45\textwidth}
\centering
\begin{tabular}{@{}l*{3}{S}@{}}
\toprule
\multirow{2}{*}{Parameter/Feature} & \multicolumn{3}{c}{atomic number scheme} \\
\cmidrule(lr){2-4}
 & {Linear} & {LASSO} & {Ridge} \\
\midrule

\rowcolor{lightgray}
\multicolumn{4}{@{}l}{\textbf{FP}} \\
Wiener & -0.5937 & -0.2067 & -0.5909 \\
RDD & -0.2688 & \zero & -0.2740 \\
TEI & 1.3610 & 0.3359 & 1.3401 \\
ECI & -0.8015 & \zero & -0.7819 \\
Harary & 1.1066 & 0.6692 & 1.1105 \\
\addlinespace[2mm]


\bottomrule
\end{tabular}
\end{minipage}
\hspace{0.02\textwidth} 
\begin{minipage}[t]{0.45\textwidth}
\centering
\begin{tabular}{@{}l*{3}{S}@{}}
\toprule
\multirow{2}{*}{Parameter/Feature} & \multicolumn{3}{c}{atomic mass scheme} \\
\cmidrule(lr){2-4}
 & {Linear} & {LASSO} & {Ridge} \\
\midrule

\rowcolor{lightgray}
\multicolumn{4}{@{}l}{\textbf{FP}} \\
Wiener & -0.5887 & -0.2067 & -0.5860\\
RDD & -0.2527 & \zero & -0.2581 \\
TEI & 1.3571 & 0.3733 & 1.3360 \\
ECI & -0.8000 & \zero & -0.7802 \\
Harary & 1.0886 & 0.6689 & 1.0926 \\
\addlinespace[2mm]


\bottomrule
\end{tabular}
\end{minipage}

\vspace{2mm}
\begin{minipage}[t]{0.45\textwidth}
\centering
\begin{tabular}{@{}l*{3}{S}@{}}
\toprule
\multirow{2}{*}{Parameter/Feature} & \multicolumn{3}{c}{atomic number scheme} \\
\cmidrule(lr){2-4}
 & {Linear} & {LASSO} & {Ridge} \\
\midrule

\rowcolor{lightgray}
\multicolumn{4}{@{}l}{\textbf{Polar}} \\
RDD & 0.4337& 0.5355 & -0.7023 \\
Gutman & -0.4050 & -0.0923 & 0.0934 \\
TEI & -0.2330 & \zero & 1.6379 \\
ECI & 0.9450 & 0.5427 & -1.0744 \\
Harary & 0.2623 & \zero & 0.8606 \\
\addlinespace[2mm]


\bottomrule
\end{tabular}
\end{minipage}
\hspace{0.02\textwidth} 
\begin{minipage}[t]{0.45\textwidth}
\centering
\begin{tabular}{@{}l*{3}{S}@{}}
\toprule
\multirow{2}{*}{Parameter/Feature} & \multicolumn{3}{c}{atomic mass scheme} \\
\cmidrule(lr){2-4}
 & {Linear} & {LASSO} & {Ridge} \\
\midrule

\rowcolor{lightgray}
\multicolumn{4}{@{}l}{\textbf{Polar}} \\
RDD & 0.4335 & 0.5373 & 0.4407 \\
Gutman & -0.4065 & -0.0949 & -0.4046 \\
TEI & -0.2440 & 0.3733 & -0.2287 \\
ECI & 0.9567 & 0.5437& 0.9403 \\
Harary & 0.2636 &\zero & 0.2556 \\
\addlinespace[2mm]

\bottomrule
\end{tabular}
\end{minipage}

\label{tab:comparison}
\end{table}

\begin{table*}
\scriptsize
\begin{minipage}[t]{0.45\textwidth}
\centering
\begin{tabular}{@{}l*{3}{S}@{}}
\toprule
\multirow{2}{*}{Parameter/Feature} & \multicolumn{3}{c}{atomic number scheme} \\
\cmidrule(lr){2-4}
 & {Linear} & {LASSO} & {Ridge} \\
\midrule

\rowcolor{lightgray}
\multicolumn{4}{@{}l}{\textbf{EV}} \\
Wiener & 2.3255 & \zero & 2.2260 \\
DD & -2.8604 & -0.2500 & -2.7586\\
RDD & 1.6706 & \zero& 0.3682 \\
TEI & 0.3123 & 0.1927 & 0.3133 \\
Harary & 0.6774 & 0.7787 & 0.7014 \\
\addlinespace[2mm]
\bottomrule
\end{tabular}
\end{minipage}
\hspace{0.02\textwidth} 
\begin{minipage}[t]{0.40\textwidth}
\centering
\begin{tabular}{@{}l*{3}{S}@{}}
\toprule
\multirow{2}{*}{Parameter/Feature} & \multicolumn{3}{c}{atomic mass scheme} \\
\cmidrule(lr){2-4}
 & {Linear} & {LASSO} & {Ridge} \\
\midrule

\rowcolor{lightgray}
\multicolumn{4}{@{}l}{\textbf{EV}} \\
Wiener & 2.3295 & \zero & 2.2296 \\
DD & -2.8574 & -0.2452 & -2.7552 \\
RDD & 0.4112 & \zero & 0.3838 \\
TEI & 0.3082 & 0.1924 & 0.3092 \\
Harary & 0.6591 & 0.8894 & 0.6831 \\
\addlinespace[2mm]
\bottomrule
\end{tabular}
\end{minipage}
\end{table*}

However, while linear models provide interpretability and computational efficiency, they may be insufficient for complex, non-linear relationships inherent in molecular structures. Therefore, non-linear prediction methods are applied in the next section. 
 
\section{Non-linear Regression Prediction}
To address the limitations of linear regression, we applied non-linear machine learning techniques, including Random Forest, XGBoost, and Neural Networks. These methods allow for the modeling of intricate dependencies between topological indices and physicochemical properties, capturing interactions that may be missed by linear approaches. By leveraging non-linear methods, we aim to improve predictive accuracy and explore the possibility of uncovering hidden patterns in the relationship between molecular structure and physical properties. The following section presents the implementation and performance analysis of these advanced machine learning models.

To optimize the Random Forest (RF) and XGBoost (XGB) models, we conducted rigorous hyperparameter tuning using grid search and cross-validation. For RF, critical parameters such as $max-depth, min-samples-split$ and $n-estimators$ were adjusted to balance the complexity and generalization of the model, prioritizing configurations that minimized overfitting (e.g. $max-depth=None$ for BP). Similarly, XGB optimization focused on parameters like $learning-rate, max-depth$, and subsample to enhance hierarchical feature interactions while avoiding excessive complexity (e.g., $max-depth=3$ for MV). In contrast, due to the limited dataset size, a shallow Neural Network (NN) architecture was implemented, comprising only one hidden layer with dropout regularization to mitigate overfitting risks. This design choice reflected the need to align model capacity with data availability—deep architectures risked memorizing noise, whereas a shallow NN preserved the ability to capture non-linear patterns without compromising stability. The optimization strategies for RF/XGB emphasized interpretability and robustness, while the NN’s simplicity underscored the trade-off between complexity and generalizability in small-data regimes.


\subsection{Comparative Analysis of Model Performance for Molecular Property Prediction}

\begin{figure}
    \centering
    \includegraphics[width=1\linewidth]{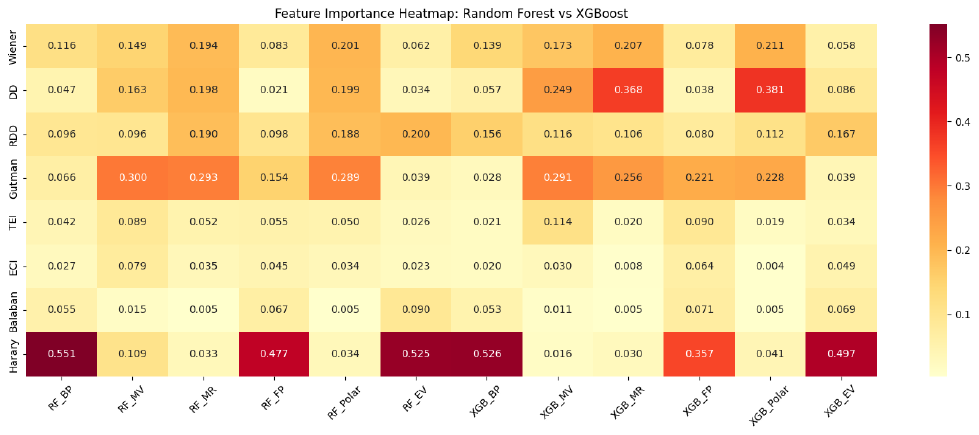}
    \vskip -12.5cm
\caption{Feature Importance Heatmap: Random Forest vs XGBoost with respect to Atomic Mass}
    \label{fig:AMRFXGB}
\end{figure}

A comparative evaluation of Random Forest (RF), XGBoost (XGB), and Neural Network (NN) models was conducted across six molecular properties—boiling point (BP), molecular volume (MV), molecular refractivity (MR), flash point (FP), polarizability (Polar), and enthalpy of vaporization (EV). The results reveal distinct performance patterns influenced by data linearity, noise levels, and model architecture.
\begin{itemize}
    \item {\bf Boiling Point (BP)}: 
For BP prediction, the NN model achieved the highest coefficient of determination (\(R^2 = 0.624 \pm 0.151\)), surpassing RF (\(R^2 = 0.51\)) and XGB (\(R^2 = 0.47\)) (Table \ref{tab:R2}). However, NN’s elevated mean absolute error (MAE = 66.98 versus 58.25 for RF) and its high variability suggest a sensitivity to noisy data and potential overfitting, despite its capability to capture nonlinear relationships. In contrast, RF’s performance appears more robust, likely due to its emphasis on high-variance linear features, such as the Harary index (feature importance = 0.551, Fig. \ref{fig:AMRFXGB}).
\item {\bf Molecular Volume (MV)}:  
For MV, both tree-based models (RF and XGB, each with \(R^2 = 0.91\)) outperformed NN (\(R^2 = 0.875 \pm 0.050\)). Notably, XGB recorded the lowest root mean square error (RMSE = 25.25 compared to NN’s 37.11). NN’s variability, particularly an RMSE of 55.92 in Fold 3, indicates its susceptibility to outliers in heterogeneous datasets. XGB’s advantage can be attributed to its effective capture of hierarchical interactions (e.g., Gutman and DD indices), which enhances its robustness in modeling nonlinear patterns (see Table \ref{tab:errorRMSE} and Fig. \ref{fig:AMRFXGB}).
\item {\bf Molecular Refractivity (MR)}:  
All models demonstrated strong consistency for MR (\(R^2 \approx 0.95\), Table \ref{tab:R2}), reflecting well-defined physicochemical patterns. However, RF exhibited marginally lower MAE (4.98 versus 6.31 for NN) and RMSE (6.79 versus 8.01 for NN), suggesting higher efficiency in stable, low-noise environments. Although NN achieved a comparable \(R^2\), its higher error metrics may indicate minor overfitting, possibly due to unnecessary complexity when modeling straightforward relationships.
\item {\bf Flash Point (FP)}:  
FP results further underscore RF’s superiority (\(R^2 = 0.67\); MAE = 33.26), outperforming both XGB (\(R^2 = 0.64\)) and NN (\(R^2 = 0.607 \pm 0.106\)) (Tables \ref{tab:R2} and \ref{tab:errorMAE}). The pronounced variability in NN’s cross-validation results (with \(R^2\) values ranging from 0.475 to 0.786) contrasts with the stable performance of RF, reinforcing the reliability of tree-based models for datasets with inherent noise and linear characteristics.

\item {\bf Polarizability (Polar)}:  
In predicting polarity, XGB and NN performed comparably (\(R^2 \approx 0.93\)–0.94), although XGB achieved a lower RMSE (3.13 versus 3.41 for NN). NN’s competitive \(R^2\) coupled with a higher MAE (2.60 compared to 2.25 for XGB) suggests that, while NN effectively captures complex interactions (e.g., DD index synergies), it may incur slight overfitting.

\item {\bf Enthalpy of Vaporization (EV)}:  
For EV, RF remained dominant (\(R^2 = 0.72\); MAE = 6.22), whereas NN lagged behind (\(R^2 = 0.636\); MAE = 9.21), further illustrating RF’s efficacy in managing linear, high-variance regimes.
\end{itemize}

Generally, RF prioritizes high-variance features, such as the Harary index for linear tasks (BP, EV), while XGB leverages non-linear interactions (e.g., the DD index in Polar predictions). Although NN is less interpretable, it complements these models by detecting latent patterns in complex datasets.
The results indicate that model selection should be aligned with the specific characteristics of the data and the analytical objectives. For tasks dominated by noise or linear relationships (e.g., BP, FP, EV), tree-based methods like RF and XGB are preferable due to their interpretability and stability. Conversely, NN may be more suitable for high-dimensional, nonlinear problems (e.g., Polar) when accompanied by adequate regularization. A hybrid framework that integrates NN’s pattern detection capabilities with the feature importance insights from tree-based models may further enhance predictive accuracy.


\begin{table}[H]
\centering
\begin{tabular}{lrrr}
\hline
 & RF ($R^2$)  & XGB ($R^2$) &  NN ($R^2$)  \\
\hline
BP      & 0.5095   & 0.4200 & 0.624 ± 0.151  \\
MV      & 0.9046   & 0.9185 & 0.875 ± 0.050\\
MR      & 0.9539 &  0.9507 & 0.951 ± 0.017 \\
FP      & 0.6798 &  0.6384 & 0.607 ± 0.106 \\
Polar   & 0.9307 &  0.9331 & 0.941 ± 0.022 \\
EV  & 0.7227 &  0.6699 & 0.636 ± 0.138 \\
\hline
\end{tabular}
\caption{Comparison $R^2$ score for RF, XGB and NN with respect to Atomic Mass}
\label{tab:R2}
\end{table}

\begin{table}[H]
\centering
\begin{tabular}{lrrr}
\hline
 &  RF (RMSE) &  XGB (RMSE)&  NN(RMSE) \\
\hline
BP      & 108.4941 &  117.9754 &  91.070  \\
MV      & 25.9555  & 23.9915  & 37.108 \\
MR      & 6.7561   & 6.9848   & 8.012 \\
FP      & 48.2114  &  51.2353  & 57.352 \\
Polar   & 3.2206   &  3.1661   & 3.409 \\
EV  &  9.2299   &  10.0702   & 11.978 \\
\hline
\end{tabular}
\caption{Comparison RMSE  for RF, XGB and NN with respect to Atomic Mass}
\label{tab:errorRMSE}
\end{table}

\begin{table}[H]
\centering
\begin{tabular}{lrrr}
\hline
 &  RF (MAE) &  XGB (MAE) & NN(MAE)  \\
\hline
BP      &  58.2028 &  70.2582 & 66.983 \\
MV      &  19.8386 &  18.3201 & 29.766 \\
MR      &  4.9935  &  5.1090  & 6.311 \\
FP      & 32.8497 &  33.8694 & 43.582 \\
Polar   &  2.1874  &  2.1462  & 2.597 \\
EV  &  6.2744  &  6.9495  & 9.213\\
\hline
\end{tabular}
\caption{Comparison MAE score for RF, XGB and NN with respect to Atomic Mass}
\label{tab:errorMAE}
\end{table}

\section{Comparative Analysis of Model Performance for a Set of Atomic Properties}

\begin{figure}
    \centering
    \includegraphics[width=0.5\linewidth]{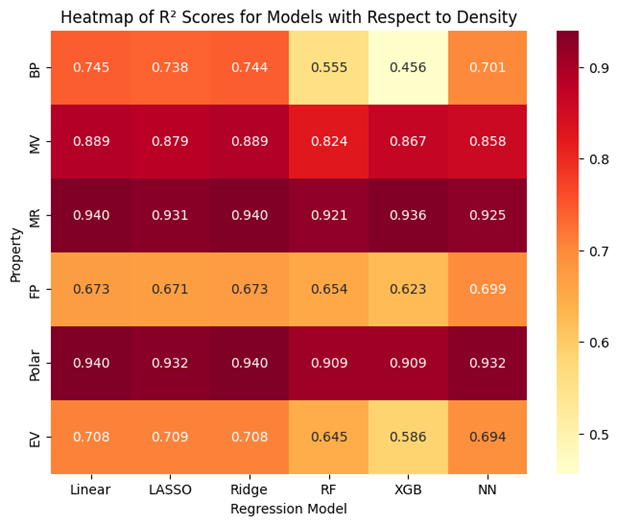}
   \caption{Heatmap of $R^2$ scores for models with respect to Density}
    \label{den}
\end{figure}

\begin{figure}
    \centering
    \includegraphics[width=0.5\linewidth]{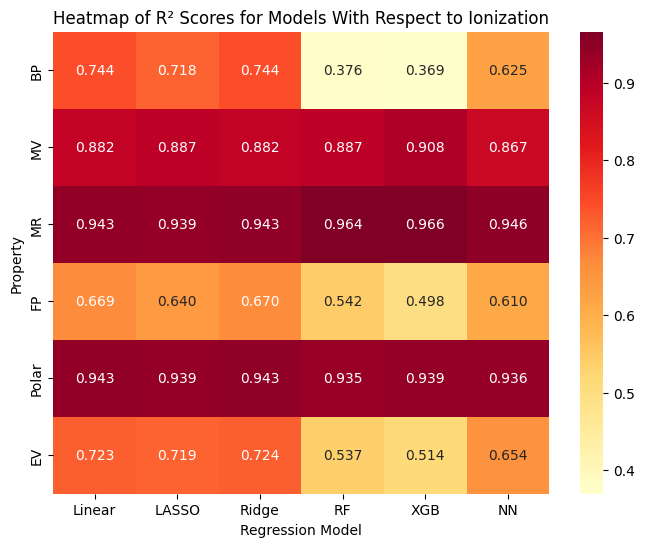}
    \caption{Heatmap of $R^2$ scores for models with respect to Ionization}
    \label{ion}
\end{figure}

\begin{figure}
    \centering
    \includegraphics[width=0.5\linewidth]{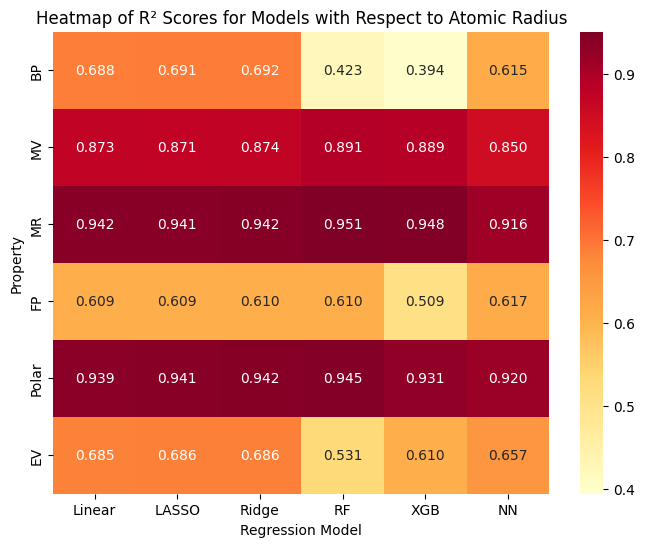}
    \caption{Heatmap of $R^2$ scores for models with respect to Atomic Radius}
    \label{AR}
\end{figure}

The predictive performance of regression models was evaluated to estimate six physicochemical properties: boiling point (BP), molar volume (MV), molar refractive index (MR), flash point (FP), polarizability (Polar), and enthalpy of vaporization (EV). Instead of focusing solely on atomic mass and atomic number, we analyzed a broader set of atomic properties, including atomic radius, atomic mass, density, ionization energy, electronegativity, and atomic number, simultaneously. Given that these atomic properties exhibit correlated influences on molecular behavior, their combined consideration provides a more comprehensive representation of atomic-level contributions to molecular descriptors. The results, summarized in Tables \ref{analAN}, \ref{analAM} and \ref{tab:R2} and Figures \ref{elc}, \ref{den},\ref{ion}  and \ref{AR}, highlight the predictive capabilities of different regression models and the impact of these atomic properties on property prediction.

\subsection{Dataset Statistical Properties and Descriptive Analysis}

To assess the statistical properties of the dataset, a comprehensive descriptive analysis was performed on the physicochemical properties. The summary statistics reveal substantial variability across different molecular properties, with BP, MV, and MR exhibiting high standard deviations (159.53, 115.28, and 39.62, respectively), indicative of a wide range of molecular behaviors. Notably, BP ranges from 133.08 to 1004.40, while MV spans from 58.80 to 811.20, underscoring significant dispersion in these properties. Additionally, skewness and kurtosis analyses show that MV (skewness: 0.8941, kurtosis: 2.2875) and MR (skewness: 0.7714, kurtosis: 1.5245) possess positively skewed distributions, whereas BP and FP display near-normal behavior.

Normality testing using the Shapiro-Wilk test further confirms deviations from normality in most physicochemical properties, with BP being the only exception, failing to reject the null hypothesis at the 0.05 significance level (p-value = 0.0532). This deviation from normality suggests that traditional parametric models may not be optimally suited to capturing the intricate relationships between topological and physicochemical properties.

Variance analysis highlights considerable dispersion in BP (variance: 25449.79) and MV (variance: 13290.05), reflecting heterogeneity in molecular structures. Moreover, the presence of outliers, particularly in MV, MR, Polar, and EV, underscores the necessity of employing robust modeling techniques to account for extreme values. Given that multiple physicochemical properties exhibit heavy-tailed distributions, standard statistical assumptions such as homoscedasticity and normality may not hold. This reinforces the need for advanced statistical methodologies and machine learning approaches that are better equipped to handle such data complexities.

\subsection{Model Selection and Predictive Accuracy}



\begin{figure}
    \centering
    \includegraphics[width=0.5\linewidth]{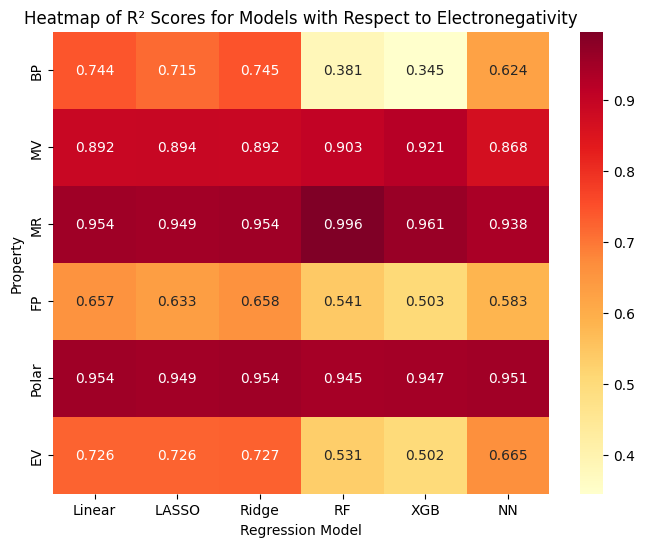}
   \caption{Heatmap of $R^2$ scores for models with respect to Electronegativity}
    \label{elc}
\end{figure}

Across all datasets, Ridge regression consistently demonstrated superior performance in predicting six target physicochemical properties, underscoring its robustness in handling collinear molecular descriptors. The highest \(R^2\) values were observed for MR and Polar across all atomic property-based analyses (\(R^2 \sim 0.95-0.96\)), indicating strong correlations with topological indices. Similarly, MV exhibited a high \(R^2\) score (\(\sim 0.9\)) in the Electronegativity dataset (Fig. \ref{elc}), reflecting reliable predictability.  

Both linear and non-linear regression models yielded similar \(R^2\) scores for MR, Polar, MV, and FP, suggesting that the relationship between atomic properties and these physicochemical characteristics is well captured regardless of model complexity. This indicates that the structural patterns influencing these properties are effectively represented through the topological indices. However, FP consistently exhibited the lowest \(R^2\) values (~0.69 in the best case), suggesting a more intricate molecular dependency that remains challenging to model even with non-linear approaches. These results highlight the importance of selecting appropriate descriptors and model architectures to optimize predictive performance across diverse physicochemical properties.

The error metrics further reinforce these findings: MR and Polar exhibited the lowest Root Mean Squared Error (RMSE) and Mean Absolute Error (MAE), confirming their strong predictability. Conversely, FP displayed the highest RMSE and MAE, aligning with its lower \(R^2\) value and greater prediction difficulty. These results highlight the varying degrees of predictability among different physicochemical properties and the critical role of atomic properties in determining the effectiveness of regression models.  

Neural Networks (NN) demonstrated superior predictive performance for boiling point (BP) compared to Random Forest (RF) and XGBoost (XGB), suggesting that NN more effectively captures the complex, non-linear relationships within thermal properties. The performance gap between these models highlights the advantages of deep learning techniques in handling intricate molecular interactions that traditional tree-based methods may not fully exploit. Additionally, in linear regression models utilizing atomic properties—electronegativity, density, and ionization—the $R^2$  values were consistently around 0.745. This moderate-to-strong correlation suggests that these fundamental atomic attributes provide meaningful contributions to property prediction, though some variance remains unexplained by a purely linear model. The results highlight the importance of feature selection and the potential need for non-linear methods to fully capture the intricate relationships between atomic properties and physicochemical characteristics.



However, EV showed notable variations across different atomic property representations. In the best case, linear regression achieved a higher $R^2$
  score (~0.726) for EV when using ionization energy, electronegativity, and atomic mass, whereas non-linear models did not show significant improvement. Additionally, RF with atomic mass yielded an $R^2$
  score close to 0.72 for EV, emphasizing the importance of molecular representation in enhancing predictive accuracy. These variations suggest that different atomic properties contribute differently to specific physicochemical characteristics, highlighting the need for tailored descriptor selection to optimize predictive performance.

\subsection{Influence of Topological Indices on Property Prediction}

The four atomic properties — atomic number, atomic mass, electronegativity, and ionization — demonstrate a consistent dependence on topological indices (TIs). By integrating both linear regression and non-linear feature importance methods, we observe a robust predictive capacity for these indices. In particular, linear regression models identify Harary, RDD, TEI, and ECI as the most influential TIs for these properties (Tables~\ref{analAN} and \ref{analAM}). Non-linear approaches, such as XGBoost and Random Forest, further underscore strong correlations of Harary and RDD with BP, FP, and EV, highlighting their relevance across various predictive frameworks (Fig. ~\ref{fig:AMRFXGB}). Additionally, Ridge Regression consistently performs well, reinforcing the central role of these indices. Meanwhile, Gutman and DD exhibit notable non-linear correlations with MV, MR, and Polar, suggesting that they complement the broader significance of Harary, RDD, TEI, and ECI.

\begin{figure}
    \centering
    \includegraphics[width=1\linewidth]{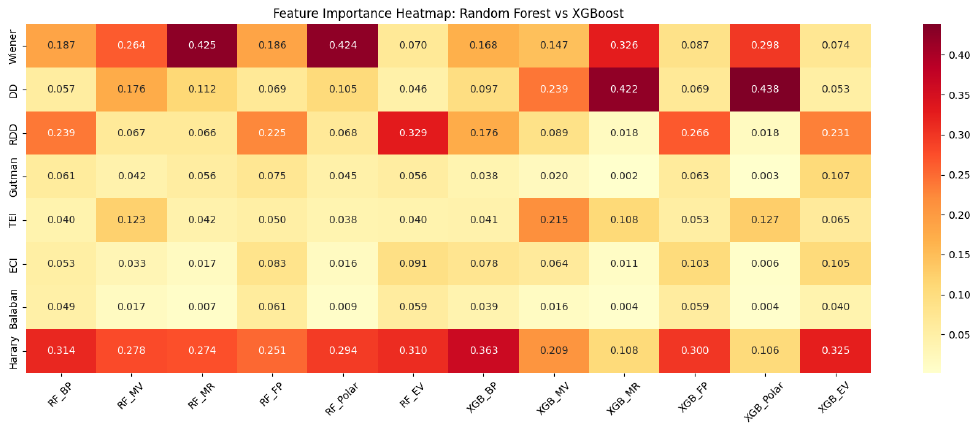}
    \vskip -13cm
    \caption{Non-linear feature important Heatmap with respect to Atomic Radius}
    \label{ARRFXGB}
\end{figure}

\begin{figure}[H]
    \centering
    \includegraphics[width=1\linewidth]{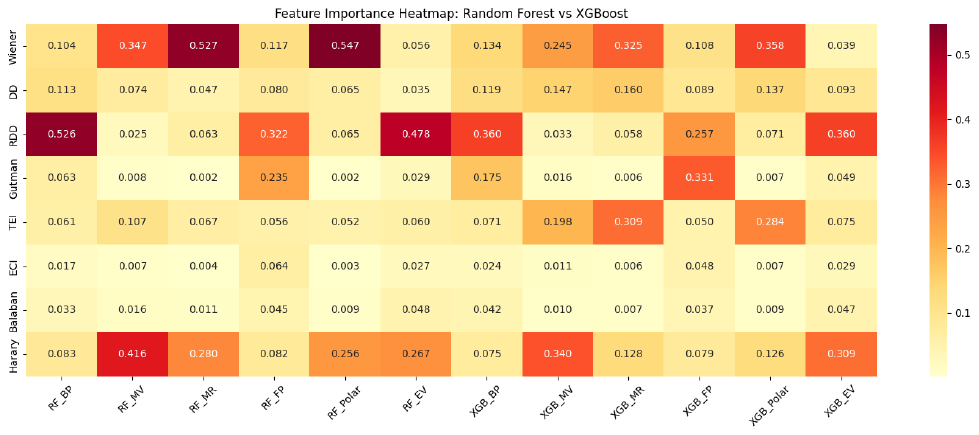}
    \vskip -13cm
    \caption{Non-linear feature important Heatmap with respect to Density}
        \label{DENRFXGB}
\end{figure}

In contrast, the correlation patterns for atomic radius and density deviate from those observed for the four atomic properties, reflecting the nuanced influence of TIs on these parameters. For atomic radius, Harary, RDD, and TEI maintain consistently high correlations across BP, MV, MR, FP, and Polar, whereas DD becomes particularly relevant for EV. This finding implies that although Harary and RDD remain broadly predictive, certain indices (e.g., DD) may be more specialized in capturing distinct aspects of atomic radius. Turning to density, RDD and Harary again prove central for BP, MR, Polar, and EV, while TEI and Wiener assume key roles for MV, MR, and EV. Notably, DD demonstrates moderate correlations with BP and FP, suggesting that the relative importance of TIs can shift depending on the targeted density sub-property. Consequently, while Harary and RDD consistently capture broad structural information pertinent to atomic radius and density, other indices such as TEI, DD, or Wiener can display strong, property-specific correlations (Tables~\ref{ARRFXGB} and \ref{DENRFXGB}). Models predicting these two parameters may therefore require a more tailored TI selection than those applied to other physicochemical properties (e.g., electronegativity or ionization), highlighting the complex interplay between molecular topology and physical attributes.

Moreover, non-linear feature importance analyses for atomic radius indicate that the Harary index remains pivotal across all sub-properties, reflecting its capacity to encode fundamental structural information. By contrast, Wiener and DD correlate strongly with MV, MR, and Polar, suggesting they capture specialized structural features, while RDD aligns more closely with BP, FP, and EV, demonstrating a distinct, property-specific influence. Together, these findings emphasize the importance of leveraging multiple topological descriptors to model atomic radius comprehensively, as each index may illuminate unique facets of molecular geometry and bonding. Likewise, in non-linear analyses for density, Harary and Wiener show particularly high correlations with MV, MR, and Polar, whereas RDD correlates most strongly with BP, FP, and EV. This partitioning of index relevance highlights the specialized roles each descriptor plays in reflecting the structural and bonding characteristics that drive density-related properties.

\begin{table}[H]
\centering
\begin{tabular}{lrrrrrr}
\hline
TIs(corrolation) &  BP &  MV & MR & FP & Polar & EV  \\
\hline
Harary      &  0.8041 &  0.9175 & 0.9481 & 0.7905 & 0.9466 & 0.8371 \\
RDD    &  0.8090 & 0.9067  &   0.9425& 0.7917 & 0.9415 & 0.8391 \\
TEI     &  0.7602  &  0.9136  & 0.9371 & 0.7754 & 0.9360 & - \\ 
DD &-&-&-&-&-& 0.7824 \\
\hline
\end{tabular}
\caption{Linear feature importance with respect to Atomic Radius }
\label{tab:errorMAE}
\end{table}

\begin{table}[H]
\centering
\begin{tabular}{lrrrrrr}
\hline
TIs(corrolation) &  BP &  MV & MR & FP & Polar & EV  \\
\hline
Harary      &  0.7722 & 0.9115  & 0.9314 & - & 0.9303 & 0.8060 \\
RDD   & 0.8388    & - &  0.9328 & 0.8365 & 0.9315 & 0.8555 \\
TEI     &  -  &  0.8796  & 0.9133 &  0.7764& 0.9120 &  \\ 
DD & 0.7562 &  - & -  & 0.7791  & - & - \\
Wiener & - & 0.8883 & - & - & - & 0.7832 \\
\hline
\end{tabular}
\caption{Linear feature importance with respect to Density }
\label{tab:corden}
\end{table}

	\section*{Declaration}
	
	\textbf{Ethical Approval:} Not applicable. \\
	
	\textbf{Competing interests:} The authors declare no any conflict of interest/competing interests.\\

	\textbf{Availability of data and materials:} All data generated or anlyzed during this study are included in this article.\\
	
	\textbf{Declaration of Generative AI and AI-assited techologies in the writin process:} During the preparation of this work the authors used ChatGPT 3.5 in order to improve readability and language of the manuscript. After using this tool/service, the authors reviewed and edited the content as needed and take full responsibility for the content of the publication.\\
	
	\textbf{Author's contributions:} All the authors [ Mohammad Javad Nadfafi Arani,  Mahsa Mirzargar] have equally contributed to this manuscript in all stages, from conceptualization to write-up of final draft.
	
	\textbf{Funding:} No funding.
	
	\section*{Acknowledgment}
	We would like to express our gratitude to the Visiting or Sabbatical Scientist Support Program for their invaluable support. Additionally, the first author acknowledges the financial support from the Scientific and Technological Research Council of Turkey (TUBITAK) under the BİDEB 2221 program.


\begin{thebibliography}{99}

    \bibitem{alz} Y. H. Mohammed,  M. Suresh and H. H. B. Jalal , Topological Indices and QSPR/QSAR Analysis of Some Drugs Being Investigated for the Treatment of Alzheimer's Disease
		Patients, \textit{Baghdad Science Journal 22}, no.1 (2025) pp.242-272. 
		https://doi.org/10.21123/bsj.2024.10866
		
		\bibitem{cordi} F.B. Farooq, N. U. H. Awan, S. Parveen, N. Idrees,
		S. Kanwal and T. A. Abdelhaleem, Topological Indices of Novel Drugs Usedin Cardiovascular Disease Treatment and Its QSPR Modeling, \textit{Journal of Chemistry}  (2022): Article ID 9749575.
		https://doi.org/10.1155/2022/9749575
		
		\bibitem{diab} S. Parveen, N. U. H. Awan, M. Mohammed,
		F. B. Farooq and N. Iqbal, Topological Indices of Novel Drugs Used in Diabetes Treatment and Their QSPR Modeling, \textit{Journal of Mathematics}  (2022): Article ID 5209329.
		https://doi.org/10.1155/2022/5209329
		
		\bibitem{sizof} X. Zhang, M. J. Saif, N. Idrees, S. Kanwal, S. Parveen, F. Saeed, QSPR Analysis of Drugs for Treatment of Schizophrenia Using Topological Indices, \textit{ACS Omega 8}  (2023): 41417--41426.
		https://pubs.acs.org/doi/10.1021/acsomega.3c05000
		
		\bibitem{Malaria} X. Zhang, H. G. G. Reddy, A. Usha, M. C. Shanmukha, M.R. Farahani and M. Alaeiyan, A study on anti-malaria drugs using degree-based topological indices through QSPR analysis, \textit{Mathematical Bioscience and Engineering} 20 (2) (2022): 3594–3609. 	
		https://doi.org/10.3934/mbe.2023167
		
		\bibitem{cyto} H. M. Nagesh, Degree-based topological indices and QSPR analysis of Cytomegalovirus drugs, \textit{International Journal of Mathematics for Industry} 2450030 (2025): https://doi.org/10.1142/S2661335224500308 
		
		\bibitem{heart} M. Hasani and M. Ghods, Topological indices and QSPR analysis of some chemical structures applied for the treatment of heart patients, \textit{Int J Quantum Chem.} (2024): 24:e27234. https://doi.org/10.1002/qua.27234
		
		\bibitem{lung} M. Arockiaraj, J. J. J. Godlin, S. Radha, T. Aziz and M. Al-harbi, Comparative study of degree, neighborhood and reverse degree based indices for drugs used in lung cancer treatment through QSPR analysis,  \textit{Scientific Reports 15} (2025): 3639. https://doi.org/10.1038/s41598-025-88044-x
		
		\bibitem{lyme} R. Huang, A. Mahboob, M. W. Rasheed, S. M. Alam and
		M. K. Siddiqui, On molecular modeling and QSPR analysis of lyme disease medicines via topological indices, \textit{Eur. Phys. J. Plus 138} (2023): article no 243. https://doi.org/10.1140/epjp/s13360-023-03867-9
		
		\bibitem{pulmo}H. Qin, M. Hussain, M. F. Hanif, M. K. Siddiqui,
		Z. Hussain and  M. A. Fiidow, On QSPR analysis of pulmonary cancer drugs using python-driven topological modeling, \textit{Scientific Reports 15} (2025): 3965. 
		https://doi.org/10.1038/s41598-025-88419-0

       \bibitem{tuber} M. Adnan, S. A. U. Bokhary ,G. Abbas and T. Iqba, Degree-Based Topological Indices and QSPR Analysis of Antituberculosis Drugs, \textit{Journal of Chemistry} (2022): Article ID 5748626. 
       https://doi.org/10.1155/2022/5748626
       
        \bibitem{viti} S. Parveen, N. U. H. Awan, F. B. Farooq, R. Fanja and Q. A. Anjum, Topological Indices of Novel Drugs Used in Autoimmune Disease Vitiligo Treatment and Its QSPR Modeling, \textit{BioMed Research International} (2022): Article ID 6045066. https://doi.org/10.1155/2022/6045066

        \bibitem{sezer} S. Sorgun and K. Birgin, "Vertex-Edge Weighted Molecular Graphs: A study on topological indices and their relevance to physicochemical properties of drugs in use cancer treatment",  \textit{Journal of Chemical Information and Modeling 65}, no. 4 (2025): pp-2093-2106. https://doi.org/10.1021/acs.jcim.4c02013

        \bibitem{huang} C. Huang, W. Gao, Y. Zheng, W. Wang, Y. Zhang, K. Liu,  Universal machine-learning algorithm for predicting adsorption performance of organic molecules based on limited data set: Importance of feature description, \textit{Science of The Total Environment} 859(1) (2023) no.160228.
    https://doi.org/10.1016/j.scitotenv.2022.160228

    \bibitem{qi} Z. Qi, S. Zhong, X. Huang, Y. Xu, H. Zhang, B. Shi, Concentration division for adsorption coefficient prediction using machine learning with Abraham descriptors: Data-splitting approach comparison and critical factors identification,\textit{Carbon} 230 (2024) no.119573.
    https://doi.org/10.1016/j.carbon.2024.119573
		
		\bibitem{book} O. Ivanciuc, T. Ivanciuc and A. T. Balaban, "Vertex-and edge-weighted molecular graphs and derived structural descriptors, Topological Indices and Related Descriptors in QSAR and QSPR", Devillers J.,Balaban, A.T. (Eds.), Taylor Francis, New York, (2018), pp. 169–220.
        https://doi.org/10.1201/9781482296945

        \bibitem{hand} J. Gasteiger, "Handbook of Chemoinformatics", Representation of Molecular Structures 4 Wiley-VCH, (2003): pp. 103-113. DOI:10.1002/9783527618279
	
		\bibitem{Wiener} I. Gutman and O.E. Polansky, "Topological Indices", \textit{Mathematical Concepts in Organic Chemistry} (1986): 123–134.https://doi.org/10.1007/978-3-642-70982-1

        \bibitem{klavzar2015} S.~Klav\v zar, M.J.~Nadjafi-Arani, Cut method: update on recent developments and equivalence of independent approaches, \textit{Curr.\ Org.\ Chem.} 19 (2015) 348--358. https://doi.org/10.2174/1385272819666141216232659

        \bibitem{klavzar2018} S.~Klav\v zar, M. J.~Nadjafi-Arani, Partition distance in graphs, \textit{J. Math. Chem.} 56(1) (2018) 69-80. https://doi.org/10.1007/s10910-017-0781-5

        \bibitem{khoda2011} H. Khodashenas, M. J. Nadjafi-Arani, A. R. Ashrafi, I. Gutman, A new proof of the Szeged-Wiener theorem, \textit{Kragujevac Journal of Mathematics} 35 (2011) 165-172.https://eid.org/2-s2.0-80051858471

        \bibitem{nad2012} M. J. Nadjafi-Arani, H. Khodashenas, A. R. Ashrafi, Relationship between edge Szeged and edge Wiener indices of graphs, \textit{Glasnik matematički} 47(1) (2012) 21-29. https://doi.org/10.3336/gm.47.1.02

        \bibitem{klavzar2014} S.~Klav\v zar, M.J.~Nadjafi-Arani, On the difference between the revised Szeged index and the Wiener index, \textit{Discrete Math.} 333 (2014) 28-34. https://doi.org/10.1016/j.disc.2014.06.006

        \bibitem{das2015} K. C. Das, I. Gutman, M. J. Nadjafi–Arani, Relations between distance–based and degree–based topological indices, \textit{Applied Mathematics and Computation} 270 (2015) 142-147. https://doi.org/10.1016/j.amc.2015.08.061
		

        \bibitem{das} K. C. Das, M. J. Nadjafi-Arani, Comparison between the Szeged index and the eccentric connectivity index, \textit{Disc. Appl. Math.} 186 (2015) 74-86. https://doi.org/10.1016/j.dam.2015.01.011
        
		
		 \bibitem{Harary} D. Plavši\'c, S. Nikoli\'c, N. Trinajsti\'c, Z. Mihali´c, On the Harary index for the characterization of chemical graphs, \textit{J. Math. Chem.} 12 (1993), 235–250. https://doi.org/10.1007/BF01164638
		
		\bibitem{balaban} A. T. Balaban, Highly discriminating distance–based topological index, \textit{Chem. Phys. Lett.} 89 (1982) 399–404. https://doi.org/10.1016/0009-2614(82)80009-2
		
		
		
		
        
        \bibitem{szr} S. Sorgun, H.Küçük and K. Birgin, Some distance-based topological indices of certain polysaccharides, Journal of Molecular Structure, 1250 (2) (2022) https://doi.org/10.1016/j.molstruc.2021.131716
		 
		\bibitem{RDD} H. Hua and S. Zhang, On the reciprocal degree distance of graphs, \textit{Discrete Appl. Math.}, 160 (2012), pp. 1152-1163. https://doi.org/10.1016/j.dam.2011.11.032
		
		
        
        \bibitem{bishop} C. M. Bishop, Pattern Recognition and Machine Learning, New York: Springer, 2006.
        
		


	
		
	\end{thebibliography}
\end{document}